\documentclass[%
reprint,
nofootinbib,
amsmath,
amssymb,
aps, 
physrev,
]{revtex4-2}
\usepackage{bm}
\usepackage{booktabs}
\usepackage{braket}
\usepackage{dcolumn}
\usepackage{graphicx}
\usepackage{hyperref}
\usepackage{mhchem}
\usepackage{siunitx}
\newcommand*\dif{\mathop{}\!\mathrm{d}}

\newcommand*\C{\mathop{}\!\text{core}}
\newcommand*\E{\mathop{}\!\text{edge}}

\newcommand*\R{\mathop{}\!\text{node}}
\newcommand*\ECR{\mathop{}\!\text{ECR}}

\newcommand*\IOL{\mathop{}\!\text{IOL}}
\begin{document}
\title{Sensitivity Analysis of Transport and Radiation in\\NeuralPlasmaODE for ITER Burning Plasmas}
\author{Zefang Liu}\email{Contact author: liuzefang@gatech.edu}
\author{Weston M. Stacey}
\affiliation{
Fusion Research Center, Georgia Institute of Technology, Atlanta, GA, USA
}
\begin{abstract}
Understanding how key physical parameters influence burning plasma behavior is critical for the reliable operation of ITER. In this work, we extend NeuralPlasmaODE, a multi-region, multi-timescale model based on neural ordinary differential equations, to perform a sensitivity analysis of transport and radiation mechanisms in ITER plasmas. Normalized sensitivities of core and edge temperatures and densities are computed with respect to transport diffusivities, electron cyclotron radiation (ECR) parameters, impurity fractions, and ion orbit loss (IOL) timescales. The analysis focuses on perturbations around a trained nominal model for the ITER inductive scenario. Results highlight the dominant influence of magnetic field strength, safety factor, and impurity content on energy confinement, while also revealing how temperature-dependent transport contributes to self-regulating behavior. These findings demonstrate the utility of NeuralPlasmaODE for predictive modeling and scenario optimization in burning plasma environments.
\end{abstract}
\maketitle
\section{Introduction}

Achieving sustained thermonuclear fusion in tokamak reactors~\cite{wesson2011tokamaks,stacey2012fusion} requires a precise understanding and control of burning plasma behavior, particularly under the high-power deuterium-tritium (D-T) conditions anticipated in ITER~\cite{aymar2002iter,green2003iter,holtkamp2007overview}. In these plasmas, interactions among energetic fusion alpha particles, electrons, and ions give rise to complex nonlinear processes, including collisional heating, radiative losses, impurity effects, and multi-region energy transport. These dynamics are strongly influenced by both global and local plasma parameters, such as magnetic field strength, safety factor, impurity concentration, and transport coefficients. Quantifying the sensitivity of plasma behavior to these parameters is essential for robust scenario design, performance optimization, and predictive control of fusion reactors.

Previous studies~\cite{wang1997simulation,green2003iter,cordey2005scaling,hill2017confinement,hill2019burn,stacey2021nodal} have proposed various models to simulate burning plasma dynamics. However, many of these approaches rely on empirical scaling laws with limited flexibility across operational regimes. To address these challenges, NeuralPlasmaODE~\cite{liu2024application,liu2024application2} was developed as a data-informed, multi-region, multi-timescale modeling framework based on neural ordinary differential equations (Neural ODEs)~\cite{chen2018neural,faraji2025machine}. This framework builds upon prior nodal modeling of tokamak plasmas~\cite{stacey2021nodal,liu2020one,liu2021multi,liu2022multi,liu2022thesis} and has demonstrated strong performance in capturing energy transport and species interactions in both DIII-D and ITER scenarios.

In this work, we extend NeuralPlasmaODE~\cite{liu2024application,liu2024application2} to conduct a comprehensive sensitivity analysis of transport and radiation effects in ITER burning plasmas. By systematically perturbing key model parameters and evaluating their normalized sensitivities, we investigate how transport diffusivities~\cite{stacey2012fusion,stacey2021nodal}, electron cyclotron radiation (ECR) parameters~\cite{albajar2001improved,albajar2009raytec}, impurity fractions~\cite{roberts1981total,dux2000neoclassical,morozov2007impurity}, and ion orbit loss (IOL) timescales~\cite{stacey2011effect,stacey2015distribution} influence core and edge densities and temperatures. The analysis focuses on a nominal inductive operation scenario in ITER, though the findings are broadly applicable to other burning plasma conditions. This study provides new insights into the dominant control parameters in reactor-scale plasmas and highlights the value of NeuralPlasmaODE as a predictive and interpretable modeling tool for future fusion experiments.

\section{Model Overview}

NeuralPlasmaODE\footnote{\url{https://github.com/zefang-liu/NeuralPlasmaODE}}~\cite{liu2024application,liu2024application2} is a multi-region, multi-timescale model that simulates burning plasma behavior in tokamaks by partitioning the plasma into spatial regions, typically core and edge, and solving coupled particle and energy balance equations using neural ordinary differential equations (Neural ODEs)~\cite{chen2018neural}. It captures essential physics including auxiliary and fusion heating, radiative and impurity losses, collisional exchanges, ion orbit loss (IOL), and inter-region transport. These processes are embedded within a differentiable framework informed by both theory and data.

The particle balance equations for deuterons, tritons, and alpha particles in the core and edge are:
\begin{align}
    \frac{\dif n_{\sigma}^{\C}}{\dif t} &= S_{\sigma,\text{ext}}^{\C} + S_{\sigma,\text{fus}}^{\C} + S_{\sigma,\text{tran}}^{\C}, \\
    \frac{\dif n_{\sigma}^{\E}}{\dif t} &= S_{\sigma,\text{ext}}^{\E} + S_{\sigma,\text{fus}}^{\E} + S_{\sigma,\text{tran}}^{\E} + S_{\sigma,\IOL}^{\E},
\end{align}
where $\sigma \in \set{\ce{D}, \ce{T}, \alpha}$ and $n_{\sigma}^{\R}$ is the particle density in region $\R \in \set{\C, \E}$. Source terms represent external fueling, fusion production, transport between regions, and edge ion orbit loss.

Energy balance equations for ion species are:
\begin{align}
    \frac{\dif U_{\sigma}^{\C}}{\dif t} &= P_{\sigma, \text{aux}}^{\C} + P_{\sigma,\text{fus}}^{\C} + Q_{\sigma}^{\C} + P_{\sigma,\text{tran}}^{\C}, \\
    \begin{split}
        \frac{\dif U_{\sigma}^{\E}}{\dif t} &= P_{\sigma, \text{aux}}^{\E} + P_{\sigma,\text{fus}}^{\E} + Q_{\sigma}^{\E} + P_{\sigma,\text{tran}}^{\E} \\
        & \quad + P_{\sigma,\IOL}^{\E},
    \end{split}
\end{align}
with $U_{\sigma}^{\R} = \frac{3}{2} n_{\sigma}^{\R} T_{\sigma}^{\R}$ as the thermal energy density. Terms include auxiliary heating, fusion heating, collisional exchange, energy transport, and IOL terms.

The energy balance equations for electrons are:
\begin{align}
    \begin{split}
        \frac{\dif U_{e}^{\C}}{\dif t} &= P_{\Omega}^{\C} + P_{e,\text{aux}}^{\C} + P_{e,\text{fus}}^{\C} - P_{R}^{\C} + Q_{e}^{\C} \\
        & \quad + P_{e,\text{tran}}^{\C},
    \end{split} \\
    \begin{split}
        \frac{\dif U_{e}^{\E}}{\dif t} &= P_{\Omega}^{\E} + P_{e,\text{aux}}^{\E} + P_{e,\text{fus}}^{\E} - P_{R}^{\E} + Q_{e}^{\E} \\
        & \quad + P_{e,\text{tran}}^{\E},
    \end{split}
\end{align}
with ohmic heating and radiative losses (including bremsstrahlung, line radiation, and electron cyclotron radiation), and the collisional energy transfer.

Transport between core and edge is governed by diffusivities defined as parametric functions of local plasma conditions:
\begin{equation}
\begin{split}
    & \frac{\chi(\rho)}{\SI{1}{m^2/s}} = \alpha_H \left( \frac{B_T}{\SI{1}{T}} \right)^{\alpha_{B}} \left( \frac{n_e(\rho)}{\SI{E19}{m^{-3}}} \right)^{\alpha_n} \\
    & \times \left( \frac{T_e(\rho)}{\SI{1}{keV}} \right)^{\alpha_T} \left( \frac{|\nabla T_e(\rho)|}{\SI{1}{keV/m}} \right)^{\alpha_{\nabla T}} q(\rho)^{\alpha_q} \kappa(\rho)^{\alpha_{\kappa}} \\
    & \times \left( \frac{M}{\SI{1}{amu}} \right)^{\alpha_M} \left( \frac{R}{\SI{1}{m}} \right)^{\alpha_{R}} \left( \frac{a}{\SI{1}{m}} \right)^{\alpha_{a}}.
\end{split}
\label{eqn:chi-scaling}
\end{equation}
In this formulation, $\alpha_H$ is a base coefficient and the exponents $\alpha_{B}, \alpha_{n}, \alpha_{T}, \alpha_{\nabla T}, \alpha_q, \alpha_{\kappa}, \alpha_M, \alpha_R, \alpha_a$ represent sensitivities to local quantities including toroidal magnetic field $B_T$, electron density $n_e$, temperature $T_e$, temperature gradient $\nabla T_e$, safety factor $q$, elongation $\kappa$, ion mass $M$, and tokamak geometry parameters $R$ and $a$. These parameters are optimized from experimental data. This study uses values aligned with ITER's inductive scenario.

\section{Sensitivity Analysis}

To evaluate how plasma densities and temperatures respond to variations in physical model parameters, we perform a normalized sensitivity analysis. For a scalar quantity $y$ and a parameter $p$, the normalized sensitivity is defined as
\begin{equation}
    S(y \,|\, p) = \frac{p}{y} \frac{\partial y}{\partial p},
\end{equation}
which quantifies the relative change in $y$ with respect to a relative change in $p$. This formulation allows comparisons across parameters with different physical units and magnitudes.

The parameters analyzed are grouped into four categories:
\begin{itemize}
    \item \textbf{Transport diffusivity parameters}: coefficients and exponents in the parametric diffusivity model~\cite{stacey2012fusion,stacey2021nodal}, capturing dependencies on magnetic field strength, density, temperature, temperature gradient, safety factor, elongation, and geometric dimensions.
    \item \textbf{Electron cyclotron radiation (ECR) parameters}: shape factors for density and temperature profiles, and the wall reflection coefficient~\cite{albajar2001improved,albajar2009raytec}, which influence radiative losses.
    \item \textbf{Impurity fractions}: concentrations of beryllium and argon impurities~\cite{roberts1981total,dux2000neoclassical,morozov2007impurity}, which affect radiation power and effective charge.
    \item \textbf{Ion orbit loss (IOL) timescales}: characteristic timescales for particle and energy loss~\cite{stacey2011effect,stacey2015distribution} due to finite orbit width effects in the edge region.
\end{itemize}

We use the trained parameters from ITER~\cite{liu2024application2} and focus on its inductive scenario 2. Each parameter is independently perturbed around its nominal value, and the resulting steady-state core and edge densities and temperatures are computed. To complement the static sensitivity analysis, we also visualize the time evolution in selected cases to examine how specific perturbations influence the dynamic behavior and stability of the burning plasma.


\begin{table*}
\caption{Normalized sensitivities to diffusivity parameters in the ITER inductive scenario 2.}
\label{tab:sensitivity-analysis-diffusivity}
\centering
\small
\begin{tabular}{ccccccccccc}
    \toprule
    Sensitivity & $ \alpha_{H} $ & $ \alpha_{B} $ & $ \alpha_{n} $ & $ \alpha_{T} $ & $ \alpha_{\nabla T} $ & $ \alpha_{q} $ & $ \alpha_{\kappa} $ & $ \alpha_{M} $ & $ \alpha_{R} $ & $ \alpha_{a} $  \\
    \midrule
    $ n_{\ce{D}}^{\C} $ & -0.0172 & 0.1098 & -0.0582 & 0.0127 & 0.0002 & -0.0569 & 0.0258 & 0.0122 & -0.0166 & -0.0082 \\
    $ n_{\alpha}^{\C} $ & -0.0729 & 0.4417 & -0.2791 & 0.0326 & -0.0252 & -0.2561 & 0.1021 & 0.0392 & -0.0953 & -0.0442 \\
    $ T_{\ce{D}}^{\C} $ & -0.1751 & 1.0922 & 0.1180 & -1.2921 & -0.3722 & -0.4172 & 0.2739 & 0.1935 & 0.0128 & -0.0266 \\
    $ T_{\alpha}^{\C} $ & -0.0199 & 0.1210 & 0.0088 & -0.1516 & -0.0461 & -0.0497 & 0.0301 & 0.0202 & -0.0022 & -0.0044 \\
    $ T_{e}^{\C} $ & -0.0006 & 0.0041 & -0.0000 & -0.0022 & -0.0011 & -0.0011 & 0.0011 & 0.0008 & 0.0004 & 0.0001 \\
    \midrule
    $ n_{\ce{D}}^{\E} $ & -0.0905 & 0.6905 & -0.5816 & -0.0191 & 0.1448 & -0.2092 & 0.1672 & 0.1317 & 0.0526 & 0.0016 \\
    $ n_{\alpha}^{\E} $ & -0.2759 & 1.9995 & -1.9186 & -0.1479 & 0.3268 & -0.7106 & 0.4773 & 0.3430 & 0.0418 & -0.0403 \\
    $ T_{\ce{D}}^{\E} $ & -0.3151 & 1.8965 & -0.1481 & -0.4806 & -0.9434 & -1.0098 & 0.5131 & 0.2013 & -0.3493 & 0.0127 \\
    $ T_{\alpha}^{\E} $ & -0.0085 & 0.0512 & -0.0033 & -0.0125 & -0.0245 & -0.0266 & 0.0137 & 0.0056 & -0.0088 & 0.0005 \\
    $ T_{e}^{\E} $ & 0.0190 & -0.1450 & 0.1225 & 0.0048 & -0.0323 & 0.0439 & -0.0351 & -0.0277 & -0.0110 & -0.0003 \\
    \bottomrule
\end{tabular}
\end{table*}

\section{Results}

Table~\ref{tab:sensitivity-analyses} summarizes the changes in steady-state core deuteron and electron temperatures in response to perturbations in key physical parameters. These include core-edge thermal diffusivities, electron cyclotron radiation (ECR) wall reflection coefficient, impurity concentrations, and ion orbit loss (IOL) timescales. We now present detailed sensitivity results and dynamic responses for each parameter group.

\begin{table}[!h]
\caption{Summary of core temperature changes in response to parameter perturbations in the ITER inductive scenario 2.}
\label{tab:sensitivity-analyses}
\centering
\begin{tabular}{cccc}
    \toprule
    Parameter & Change & $ \Delta T_{\ce{D}}^{\C} (\si{keV}) $ & $ \Delta T_{e}^{\C} (\si{keV}) $ \\
    \midrule
    Core-edge thermal & $ \times $0.5  & +1.184 & +2.038 \\
    diffusivities $ \chi_{\sigma}^{\C} $ & $ \times $2.0 & -1.159 & -1.751 \\
    \midrule
    Wall reflection coeff. & -0.1 & -0.002 & -0.025 \\
    of ECR $ r = 0.8 $ & +0.1 & +0.002 & +0.033 \\
    \midrule
    Beryllium fraction & -1.0\% & -0.063 & +0.059 \\
    $ f_{\ce{Be}} = 2.0\% $ & +1.0\% & +0.058 & -0.072 \\
    \midrule
    Argon fraction & -0.04\% & -0.005 & +0.165 \\
    $ f_{\ce{Ar}} = 0.12\% $ & +0.04\% & -0.008 & -0.184 \\
    \midrule
    IOL timescales & $ \times $0.1 & -0.000 & -0.000 \\
    $ \tau_{p, \sigma, \IOL}^{\E} $ and $ \tau_{E, \sigma, \IOL}^{\E} $ & $ \times $10.0 & +0.000 & +0.000 \\
    \bottomrule
\end{tabular}
\end{table}

\subsection{Transport Diffusivity Sensitivities}

We first evaluate the impact of transport diffusivity parameters on core and edge plasma conditions. Normalized sensitivities with respect to the diffusivity scaling coefficients are listed in Table~\ref{tab:sensitivity-analysis-diffusivity}. Several key trends are evident:
\begin{itemize}
    \item Alpha particle densities show greater sensitivity to diffusivity parameters than deuteron densities in both core and edge regions.
    \item Core deuteron density is particularly sensitive to magnetic field, density, and safety factor exponents.
    \item Among temperatures, core deuterons are most responsive to parameter changes, especially those related to temperature, magnetic field, and safety factor.
\end{itemize}
These results highlight the magnetic field and safety factor as critical control parameters for modulating inter-region transport and maintaining plasma stability.

\begin{figure}[!h]
\centering
\includegraphics[width=.9\linewidth]{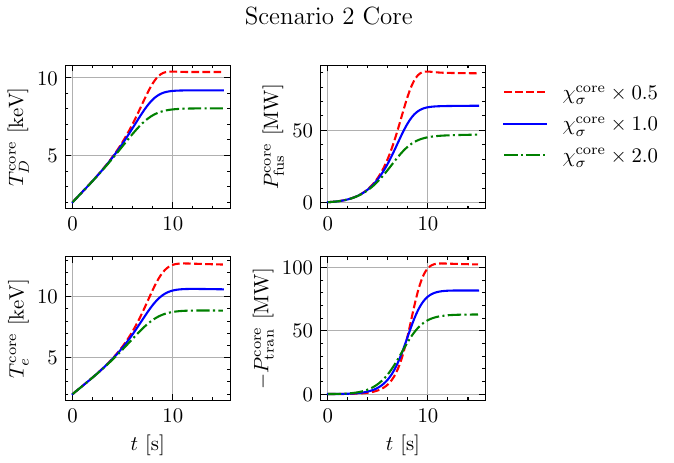}
\caption{Effect of perturbing core-edge thermal diffusivity on plasma evolution in the ITER inductive scenario 2.}
\label{fig:iter-2-sens-chi-core}
\end{figure}

To illustrate dynamic effects, we perturb the core-edge thermal diffusivities $\chi_{\sigma}^{\C}$ (for $\sigma \in \set{\ce{D}, \alpha, e}$) by factors of 0.5 and 2.0. As shown in Figure~\ref{fig:iter-2-sens-chi-core}, reducing these diffusivities limits outward energy transport, leading to higher core temperatures and increased fusion power. However, because the diffusivity model includes strong temperature and gradient dependencies, heat transport rises with core temperature, which helps offset further heating and prevents runaway behavior. The resulting changes in steady-state temperatures are summarized in Table~\ref{tab:sensitivity-analyses}.

\subsection{ECR Parameter Sensitivities}

We next analyze sensitivities related to electron cyclotron radiation (ECR). Table~\ref{tab:sensitivity-analysis-ecr} shows the normalized sensitivities of core and edge electron temperatures to four key parameters~\cite{albajar2001improved}: the density shape exponent $\alpha_{n}$, the temperature shape exponents $\alpha_{T}$ and $\beta_{T}$, and the wall reflection coefficient $r$. Among these, the wall reflection coefficient $r$ and temperature shape parameter $\beta_{T}$ exhibit the strongest influence on electron temperatures. These findings highlight the importance of accurately modeling wall reflectivity, which remains uncertain for ITER and affects how core ECR losses couple to edge and wall regions.

\begin{table}[!h]
\caption{Normalized sensitivities of electron temperatures to ECR parameters in the ITER inductive scenario 2.}
\label{tab:sensitivity-analysis-ecr}
\centering
\begin{tabular}{ccccc}
    \toprule
    Sensitivity & $ \alpha_{n} $ & $ \alpha_{T} $ & $ \beta_{T} $ & $ r $ \\
    \midrule
    $ T_{e}^{\C} $ & -0.0001 & 0.0100 & -0.0135 & 0.0212 \\
    $ T_{e}^{\E} $ & -0.0001 & 0.0098 & -0.0132 & 0.0208 \\
    \bottomrule
\end{tabular}
\end{table}

\begin{figure}[!h]
\centering
\includegraphics[width=.9\linewidth]{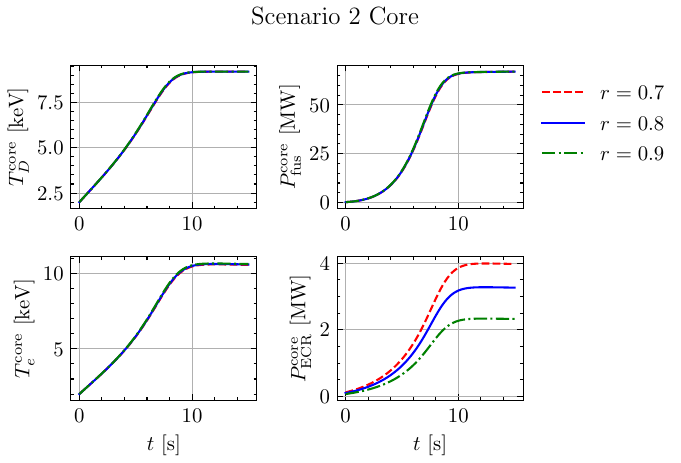}
\caption{Effect of wall reflection coefficient $r$ on ECR loss in the core region for the ITER inductive scenario 2.}
\label{fig:iter-2-sens-ecr-r}
\end{figure}

To visualize dynamic effects, Figure~\ref{fig:iter-2-sens-ecr-r} shows the time evolution of core ECR power loss $P_{\ECR}^{\C}$ for different values of $r$, with $r = 0.8$ as the nominal case. As expected, increasing $r$ reduces radiative losses by reflecting more power back into the plasma. However, due to the relatively low electron temperatures in this scenario, the impact on overall energy balance and temperature profiles is modest. Corresponding changes in core temperatures are summarized in Table~\ref{tab:sensitivity-analyses}.

\subsection{Impurity Fraction Sensitivities}

We next examine the impact of impurity concentrations on plasma temperatures. Table~\ref{tab:sensitivity-analysis-impurity} shows normalized sensitivities of electron temperatures to variations in beryllium and argon impurity fractions~\cite{roberts1981total,dux2000neoclassical,morozov2007impurity}. Negative values indicate that increasing the impurity content reduces electron temperature, primarily due to enhanced radiative losses. Argon exhibits a notably stronger effect than beryllium, consistent with its higher radiative power at ITER-relevant temperatures.

\begin{table}[!h]
\caption{Normalized sensitivities of electron temperatures to impurity fractions in the ITER inductive scenario 2.}
\label{tab:sensitivity-analysis-impurity}
\centering
\begin{tabular}{ccc}
    \toprule
    Sensitivity & $ f_{\ce{Be}} $ & $ f_{\ce{Ar}} $ \\
    \midrule
    $ T_{e}^{\C} $ & -0.0120 & -0.0469 \\
    $ T_{e}^{\E} $ & -0.0184 & -0.0693 \\
    \bottomrule
\end{tabular}
\end{table}

\begin{figure}[!h]
\centering
\includegraphics[width=.9\linewidth]{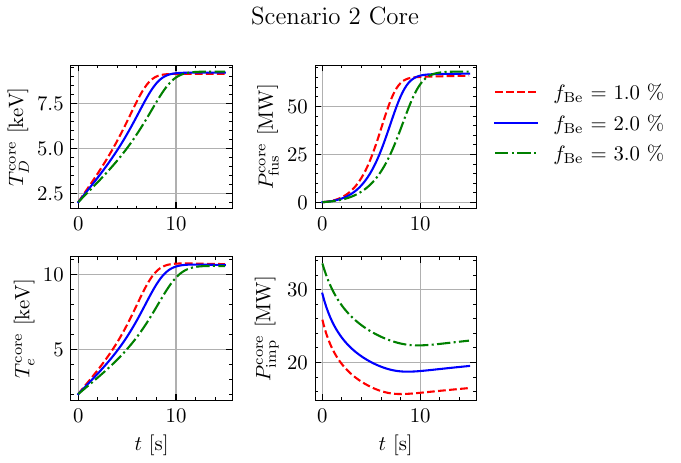}
\caption{Effect of beryllium impurity fraction on core and edge temperatures in the ITER inductive scenario 2.}
\label{fig:iter-2-sens-be}
\end{figure}

\begin{figure}[!h]
\centering
\includegraphics[width=.9\linewidth]{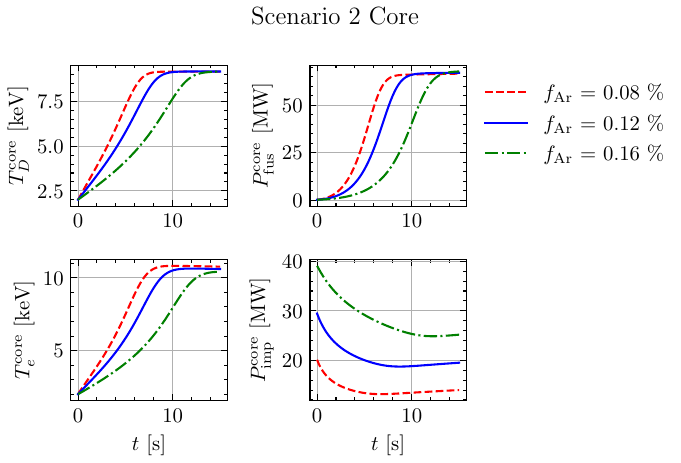}
\caption{Effect of argon impurity fraction on core and edge temperatures in the ITER inductive scenario 2.}
\label{fig:iter-2-sens-ar}
\end{figure}

Figures~\ref{fig:iter-2-sens-be} and~\ref{fig:iter-2-sens-ar} show the time evolution of deuteron and electron temperatures under perturbed impurity fractions. In both cases, increasing impurity content leads to slower temperature rise due to higher impurity radiation. However, as core electron temperatures increase over time, the radiative power of impurities diminishes, resulting in relatively modest differences in steady-state temperatures. Corresponding changes are reported in Table~\ref{tab:sensitivity-analyses}.

\subsection{Ion Orbit Loss Timescale Sensitivities}

We conclude with a sensitivity analysis of ion orbit loss (IOL)~\cite{stacey2011effect,stacey2013effect,stacey2015distribution,stacey2016inclusion,stacey2018dependence,wilks2016improvements,wilks2016calculation} timescales. Table~\ref{tab:sensitivity-analysis-iol} shows the normalized sensitivities of edge densities and temperatures to the characteristic timescales for particle and energy loss due to IOL. The results indicate negligible sensitivity in all quantities, with values close to zero. This outcome reflects the relatively small IOL loss power in the ITER edge region under typical conditions~\cite{hill2019burn}. Because IOL has limited influence on the core region and minimal impact on steady-state profiles, further dynamic simulations with varied IOL timescales are not necessary in this case.

\begin{table}[!h]
\caption{Normalized sensitivities of edge quantities to IOL timescales in the ITER inductive scenario 2.}
\label{tab:sensitivity-analysis-iol}
\centering
\begin{tabular}{ccc}
    \toprule
    Sensitivity & $ \tau_{p,\IOL}^{\E} $ & $ \tau_{e,\IOL}^{\E} $ \\
    \midrule
    $ n_{\ce{D}}^{\E} $ & \SI{-5.41e-9}{} & \SI{4.33e-9}{} \\
    $ n_{\alpha}^{\E} $ & \SI{3.48e-6}{} & \SI{8.98e-8}{} \\
    $ T_{\ce{D}}^{\E} $ & \SI{-3.04e-8}{} & \SI{5.79e-7}{} \\
    $ T_{\alpha}^{\E} $ & \SI{-1.78e-7}{} & \SI{1.50e-6}{} \\
    \bottomrule
\end{tabular}
\end{table}

\section{Discussion}

This analysis identifies transport diffusivity parameters, particularly those related to magnetic field strength and safety factor, as the most influential factors in determining core and edge behavior in ITER plasmas. Their impact on inter-regional energy and particle transport makes them critical for scenario control. The temperature dependence of diffusivities introduces a self-regulating feedback that contributes to system stability and mitigates the risk of runaway heating under parameter perturbations.

Impurity fractions, especially argon, also significantly affect radiative losses and reduce electron temperature, underscoring the importance of impurity control. ECR parameters show weaker overall influence, but uncertainties in wall reflection coefficients may still impact edge power balance. Ion orbit loss has minimal effect in this scenario, indicating that it plays a limited role under ITER-like conditions, though it may still be important in other devices or edge configurations.

\section{Conclusion}

This study presents a sensitivity analysis of transport, radiation, and impurity parameters using NeuralPlasmaODE for ITER burning plasmas. Results highlight the dominant influence of transport physics, particularly magnetic field strength, safety factor, and temperature gradients, on core behavior. Impurity fractions, especially argon, also impact electron temperatures through radiative losses. These findings demonstrate the utility of NeuralPlasmaODE in identifying key control parameters and support its application to scenario optimization and predictive modeling in future fusion experiments.

\bibliography{apssamp}
\end{document}